\documentclass[mlmain]{jmlr}

\usepackage{longtable}

\usepackage{booktabs}
\usepackage[load-configurations=version-1]{siunitx} 

\theorembodyfont{\upshape}
\theoremheaderfont{\scshape}
\theorempostheader{:}
\theoremsep{\newline}

\jmlrvolume{}
\firstpageno{1}

\title[Short Title]{Multiagent Simulators for Social Networks}

  \author{\Name{Aditya Surve\nametag{\thanks{Equal Contribution}}} \Email{ssaditya2002@gmail.com}\\
  \Name{Archit Rathod$^{*}$} \Email{architrathod77@gmail.com}\\
  \Name{Mokshit Surana$^{*}$} \Email{mokshitsurana3110@gmail.com}\\
  \Name{Gautam Malpani$^{*}$} \Email{gautammalpani33@gmail.com}\\
  \Name{Aneesh Shamraj$^{*}$} \Email{aneesh.shamraj@gmail.com}\\
  \Name{Sainath Reddy Sankepally$^{*}$} \Email{sankepallysainathreddy@gmail.com}\\
  \Name{Raghav Jain$^{*}$} \Email{raghavjain106@gmail.com}\\
  \Name{Swapneel S Mehta$^{*}$} \Email{swapneelsmehta@gmail.com}\\
  \addr SimPPL}

\begin{document}

\maketitle

\begin{abstract}
Multiagent social network simulations are an avenue that can bridge the communication gap between the public and private platforms in order to develop solutions to a complex array of issues relating to online safety.
While there are significant challenges relating to the scale of multiagent simulations, efficient learning from observational and interventional data to accurately model micro and macro-level emergent effects, there are equally promising opportunities not least with the advent of large language models that provide an expressive approximation of user behavior.
In this position paper, we review prior art relating to social network simulation, highlighting challenges and opportunities for future work exploring multiagent security using agent-based models of social networks.

\end{abstract}
\section{Introduction}
\label{sec:intro}


Simulators have become integral in various industries \cite{ottogalli2019flexible} offering virtual environments for training, testing, and experimentation \cite{bousquet1999environmental} . 
Their utility spans aviation \cite{sun2022large},\cite{xiong2022digital}, \cite{bernard2022digital}, military  \cite{kessels2021added},\cite{sattler2020simulation} healthcare \cite{kononowicz2019virtual},\cite{kumar2020novel},\cite{croatti2020integration} and entertainment\cite{woolworth2019acoustics}.
Notably, simulators provide a secure and controlled space for activities, aiding in cost reduction, safety enhancement, and efficiency improvement \cite{green2016improving}. 
They contribute to evaluating and optimizing designs before implementation, saving both time and 
 money\cite{makransky2019equivalence}.
Social networks are complex information-sharing systems that have become digital information highways \cite{borgatti2018analyzing},\cite{van2009impact}.
There have been various attempts to simulate the spread of various types of information on social media using multiagent simulators of user behavior \cite{zhou2020innovation},\cite{massaguer2006multi},\cite{zhao2015simnest},\cite{prike2023source},\cite{sakas2019modeling},\cite{dalayli2020representation}.
Multiagent simulators make it possible for external researchers to develop experiments that can be run in the virtual testbed it provides \cite{cardoso2021review}.\\
The rest of the paper is structured as follows. We first discuss the utility of simulators in social networks in Section 2. We then discuss the application of simulators across different industries in Section 3. In Section 4, we provide past work and efforts on simulating social networks. In Section 5, we provide a review of recent work on LLMs and simulations. In
Section 6, we discuss the limitations and open challenges of simulators which is followed by possibilities for future
work in Section 7 and conclude our paper in Section 8.

\section{Prior Work in Simulating Social Networks}

Social network simulations have become an integral part of understanding and predicting user behavior on platforms like Facebook, Google, and Twitter. Agent-based simulations with autonomous agents imitate real user actions. Simulators allow safe experimentation with potential changes, ensuring they don't impact real users.

As per \cite{ahlgren2020wes}, Facebook utilizes a platform called WW (Web-Enabled Simulation) to simulate user interactions and social behaviors within a parallel version of its platform. 
WW employs autonomous software agents or "bots" programmed to imitate real user actions, such as posting, messaging, and making connections. 
These bots are trained using anonymized logs of real user activity data from Facebook to make their behaviors realistic and statistically match real user statistics.
Similarly, Google built RecSim to model the societal effects of recommender systems \cite{ie2019recsim, mladenov2021recsim}. 
RecSim allows configuring agents to represent various types of users, content providers, and other participants in the recommender ecosystem.
Twitter also employs a reinforcement learning over agent models in order to optimize user engagement through push notifications \cite{o2022should}. 
 As a practical example of the real-world value drawn from simulators, their system manages to successfully maximize long-term user satisfaction by studying user responses to push notifications. 
 
 There have been other prior attempts at building expressive forward simulators.
 HashKat \cite{ryczko2017hashkat} is a dynamic network simulation tool designed to model the growth of information propagation through an online social network. 
 NetSim \cite{stadtfeld2013package} is an R package that allows for the simulation of the co-evolution of social networks and individual attributes.
 There is a history of common challenges associated with simulating social networks.

\section{Motivations to Simulate Social Networks}

This paper dives into how simulators are used as tools for modeling information dissemination in social networks, improving our ability to understand of a variety of complex issues on social networks. 

\subsection{Modeling Algorithmic Effects}
Social media algorithms used to designed to maximize engagement, which tended to amplify human biases towards learning from prestigious, ingroup, emotional, and moral (PRIME) \cite{brady2023algorithm}.
This can promote misinformation and polarization. 
Algorithms prioritize engagement over accuracy or truth, leading to the rapid spread of extreme, controversial, or false content. 
Users often find themselves in "filter bubbles" and "echo chambers," reinforcing their existing views and distorting their perception of group opinions. 
Users often find themselves in "filter bubbles" and "echo chambers," reinforcing their existing views and distorting their perception of group opinions. 
The concept of "wisdom of crowds" is compromised by online echo chambers and the presence of fake accounts, bots, and orchestrated networks that manipulate engagement signals \cite{saurwein2021automated}.

\subsection{Modeling Policy Interventions}

It is challenging to predict the multifaceted effects of policies on social media.
Multiagent simulators offer a virtual testbed for prototyping policy outcomes with the caveat that they don't necessarily reflect all the possible outcomes in perfect accord with the real-world.
However, it is useful to be able to model the relative effects of different types of policies prior to their deployent since that permits us to evaluate various policies in the same environment.
Simulators can provide a testbed to prototype interventional effects and model how they might influence the system in desirable and undesirable ways.\\

\textbf{User Experience and Exposure to Information}: The addictive nature \cite{pellegrino2022research} of social media poses challenges to individuals' well-being. 
Persuasive design techniques contribute to stress, anxiety, and decreased productivity.
Filter bubbles and information overload hinder diverse perspectives and contribute to the spread of fake news. 
Designing healthy user experiences is crucial, emphasizing user control and breaks in user flow.

\subsection{Security Testing}

\textbf{Algorithmic Vulnerability Detection}: Algorithms used by social media platforms can replicate and amplify human biases, resulting in outcomes that are less favorable to certain groups. 
Testing strategies include auditing algorithms, diverse design teams, and user feedback. 
Challenges involve trade-offs between fairness and accuracy, privacy regulations, and the need for algorithmic literacy.\\

\textbf{Social Exploitation by Coordinated Networks (CIB)}: Social bots include, automated accounts imitating human behavior, manipulate public opinion by spreading fake news and divisive content \cite{zhang2023social}. 
Detection and counteracting social bots is challenging due to their sophistication.
\cite{orabi2020detection} shows that social bots evolve rapidly to evade detection, presenting an "everlasting cat and mouse game" between bot creators and detection methods.
Information operations by coordinated networks highlight the global impact of disinformation on social networks.\\

\textbf{Polarization}: Selective exposure to attitude-confirming information exacerbates confirmation bias and polarizes opinions. 
Tolerance among users plays a role in mitigating polarization. 
Information propagation is influenced by the structure of social networks and user activity patterns \cite{haque2023understanding}.\\

\textbf{Information Propagation}: The structure of social networks impacts how information spreads. User activity patterns follow a power law distribution, with active users playing a significant role in information propagation. 
Understanding information spread is crucial for better recommendations and identification of manipulation.

\subsection{Benchmarking Recommender Systems}
Simulators can be used to model how information spreads on social networks and to examine the effects of recommender systems on the virtual sharing ecosystem when it comes to benchmarking recommendation systems. \cite{stavinova2022synthetic} demonstrates one of the utilities of simulators which is to test the limits of existing recommender systems of different types (including Reinforcement Learning ones) and to complex user preference formation.

There are several available online recommendation systems that can be used as examples to examine the effects of recommender systems on the virtual information-sharing ecosystem. One example is this \textsf{repository}.\footnote{See
\url{https://github.com/recommenders-team/recommenders}}, which contains different sophisticated recommendation systems available for simulation environments. The repository provides baselines and reproducible code for standard recommendation techniques, and the datasets available in the repository could serve as a starting point to generate simulated user-item interactions and feedback. Another example is RecSim \cite{ie2019recsim}, a configurable platform for authoring simulation environments to facilitate the study of RL algorithms in recommender systems.

\section{Simulators are Employed Across Industries}

Simulators find widespread utility across industries. 
In aviation, flight simulators, ranging from basic desktop models to complex full-motion systems, are crucial for pilot training \cite{cross2022using},\cite{saastamoinen2021usefulness},\cite{trinon2019immersive}. 
Healthcare relies on simulators for teaching medical workers and replicating complex medical procedures, spanning from fundamental task trainers to advanced patient simulators mimicking human physiology 
 \cite{rosen2012situ} ,\cite{so2019simulation}. 
Space exploration benefits from simulations for spacecraft design and mission scenario testing, allowing for development and optimization before actual missions\cite{nichols2023space},\cite{aslan20195g},\cite{liu2022space}.

Beyond aviation and healthcare, simulations play pivotal roles in cybersecurity\cite{armenia2021dynamic}, \cite{le2020gridattacksim}, economics\cite{bremer2019globus}, climate science \cite{politi2020sensitivity},\cite{rasmussen2020changes} and nuclear power plant design \cite{vajpayee2020dynamic},\cite{el2019system},\cite{wu2019development} .
These virtual environments assist in modeling, testing safety measures, and simulating emergency situations.
Illustrative examples include the 
\cite{wang2023drive} - a diffusion based simulator driven by real world autonomous driving data, \textsf{CERN's computational psychometrics}\cite{cipresso2015modeling} for real-world and virtual behavior integration, and Siemens' use of TECNOMATIX \citep{bangsow2020tecnomatix} for optimizing production systems and logistics processes. 
These applications showcase the diverse and impactful uses of simulators across various domains, demonstrating their potential in understanding and modeling complex scenarios.

\subsection{Examples}

Simulators provide a virtual testbed for prototyping various policies as interventions upon the world model underlying a social network.
They allow us to encode our understanding of the rules underpinning social interactions, incorporate design affordances from the real world, and create an arbitrarily complex model of information propagation on social networks. 
Simulators can leverage digital trace data to calibrate their parameters,attempting to provide an accurate representative model of reality.
It also allows experimentation in scenarios where there might be ethical challenges deploying randomized control trials.
For example, TACIT by \cite{neumann2023does}, enhance fact-checking models and assess their impact on reducing inequalities among online communities. 
It sparked ethical discussions on prioritizing equity in AI-driven fact-checking. 

\textbf{Algorithmic Auditing}: Regulators need detailed evidence on platforms' policies, processes, and outcomes related to misinformation. 
Algorithm auditing requires a multidisciplinary skill-set and granular data on misinformation spread. 
Modeling effects of algorithms on social networks can provide insights on balancing societal impacts and technical performance as AI moves from research into real-world applications\footnote{ \url{https://hbr.org/2018/11/why-we-need-to-audit-algorithms}}.

\textbf{Testing User Experience (UX) on Social Media}: Simulations \cite{ahlgren2020wes} in a social network can assess and predict the type of content users might be exposed to.The open source Misinformation Game simulator \cite{butler2023mis} provides a flexible and customizable platform for conducting controlled experiments on factors influencing online misinformation propagation and beliefs.
By emulating recommendation systems and user interactions, simulations can reveal issues like filter bubbles, echo chambers, and exposure to harmful content.

\textbf{Testing Security}: Red teaming and blue teaming simulations \cite{seker2018concept} assess and enhance security and privacy measures on social media platforms. 
These simulations enable proactive identification and mitigation of risks, strengthening the overall security infrastructure.

\section{Large Language Model based Agents and Generative Social Science}

The recent advancements in natural language processing and machine learning viz. LLMs has given new life to the idea of 'agent'-based models and spurred interest in generative social simulations.
Large Language Models (LLMs) are increasingly employed in generative social science, particularly in agent-based modeling.
\footnote{There is frequently an overloading of the term 'agents' where it may reference an independently operating LLM; in this context an LLM might represent a user's activity independently in a complex system.}
For example, the approach by \cite{shaikh2023rehearsal} aims to explain macro-level social phenomena by simulating interactions among individual agents following simple rules. 
The goal is to grow macro structures from micro foundations, emphasizing dynamic processes and the sufficiency of micro-level rules. 
To enhance realism, models are embedded with empirical data about agents and environments. 
Agent-based modeling is instrumental in studying diverse social phenomena, from disease spread to urban development, by unraveling complex systems' emergence from individual agent interactions. 
Integrating LLMs into this framework, as seen in projects, like Stanford's generative agents by \cite{park2023generative}, aims to introduce human-like behaviors and natural language communication in simulated environments, modeling on information diffusion, relationship formation, and coordination in these societies.

\section{Limitations and Open Challenges}

The limitations of simulations span a range of challenges. 
The dependence of simulation outcomes on specific parameter values and internal model structures necessitates sensitivity analysis for a nuanced understanding of variability. 
Additionally, the inherent complexity of model specifications often limits the transparency of simulation results, hindering a comprehensive grasp of agent trajectories and behaviors.
With bespoke simulators for the same task and a lack of transparency into the design of complex simulation procedures, reproducibility becomes challenging due to the absence of standardized procedures and model sharing, impeding knowledge accumulation across studies. 
The incorporation of social networks introduces complexity, requiring careful calibration with empirical data and risking biases towards replicating existing conditions. 
Moreover, the abstraction of human behavior in models overlooks crucial psychological and social nuances. 
Computational power and time constraints further limit the size and complexity of modeled networks. 
Finally, while achieving macro-level congruence with real-world data is a common goal, it does not guarantee accuracy at the micro-level, posing challenges in validating models against real-world data as per \cite{manzo2014potentialities}.
Additionally, there is ongoing research into LLM-human hybrid models, exemplified by Facebook's Cicero AI\cite{meta2022human}.
The ideas behind integrating LLMs into agent-based modeling extends beyond simulation, towards high-fidelity real-world experimentation introducing new challenges and opportunities at the intersection of artificial intelligence and social science.

\subsection{Open Challenges}
\label{sec:floats}

While multi-agent simulators are a promising approach for studying information for studying information propagation on social networks, current techniques have significant limitations. 
Even the most advanced simulators have significant gaps in their capabilities when compared to the complexity of real-world social platforms.

\subsubsection{Inference at Scale}

Inferring accurate simulation parameters from real-world social network data is extremely challenging, 
especially at the massive scales of modern platforms.
Each social media site has a unique architecture and focuses on different types of user interactions and content. 
For example, Twitter emphasizes short messages, broadcasting and news, 
while Instagram centers on the visual photo and video sharing. 
The core algorithms driving each platform are opaque and keep changing.

Computational social science techniques aim to improve large-scale inference by combining machine learning with insights from disciplines such as sociology, psychology, and communications theory. For example, research into cognitive biases that influence how users spread misinformation. Even with advances in big data and artificial intelligence, capturing every aspect of human behaviour remains difficult.

\subsubsection{Data Limitations}

While simulations rely on real-world data for inference and evaluation, comprehensive social media data is increasingly difficult for researchers to access. 
Platforms like Facebook and Twitter have become more restrictive in sharing data, due to concerns around privacy, ethics and potential misconduct.
For instance, in an interview by \textsf{Undark}.\footnote{See
\url{https://undark.org/2022/04/18/why-researchers-want-broader-access-to-social-media-data/}}, Meta representatives said that common researcher practices like web scraping or third-party APIs can now lead to accounts being blocked or banned if done without permissions. 
Even when data is granted, it is often limited in scope or heavily sampled across several channels.
This makes collecting large, unbiased datasets to train accurate simulations acutely challenging.
The study conducted by \citet{liu2016detecting}, mentioned how they used web crawling for their study but it was difficult to adapt it to simulations for large-scale data.
In summary, expanding platform restrictions on data access increasingly hinder the simulation research needed to understand and improve social media. 
Mechanisms to enable responsible data sharing with researchers are needed.
As per \cite{10.3389/fenvs.2015.00063}, Social media provides a wealth of real user conversation data that could help in conversational science research. 
But access is restricted so responsible data sharing under ethics would enable leveraging these conversations to advance conversational agents. With proper safeguards, social media data presents opportunities to develop dialogue systems grounded in natural human exchanges. Finding the balance between privacy protection and research access remains challenging but important for progress in conversational AI.
\cite{kapoor2018advances} also highlights the value of social media data, showing how user-generated content on these platforms provides a rich source of natural conversations and social interactions that could inform research across diverse fields, including information systems.

\subsubsection{Modeling Algorithmic Effects}
Algorithms play a crucial role when it comes to social media recommendation systems in delivering appropriate content to users. 
However, the intricacy of these algorithms can be challenging to model in simulations. 
We can take a look at social media platforms like Twitter, TikTok, Netflix and YouTube as general examples of a complex recommendation system.
TikTok's recommendation system, Monolith, is a real-time recommendation system by \cite{liu2022monolith} that incorporates data structures such as collision-less embedding tables with distributed architectures for training and serving. The concurrent data flows, timing-sensitive operations, failure handling, and sheer data volumes presented in the paper are non-trivial to model.

YouTube represents one of the largest scale and most sophisticated recommendation systems in existence as shown in \cite{paul45530}. 
The recommendation system at  
\textsf{Twitter}\footnote{
\url{https://blog.twitter.com/engineering/en_us/topics/open-source/2023/twitter-recommendation-algorithm}}
is composed of many interconnected services and jobs, which aims to distill roughly 500 million tweets posted daily down to a small number of top tweets that ultimately show up on a user's "For You" section.
On the other hand, the recommender system at Netflix, described in \cite{gomez2015netflix}, is not just one algorithm but rather a variety of algorithms that collectively define the Netflix experience. 

The paper by \cite{gao2022causal} is a survey of the literature on causal inference-based recommendation, that aims to enhance recommender systems by utilizing causal inference to extract causality from data.
To simulate algorithms used in this paper, we would require modeling complex causal relationships, accounting for potential confounders and biases, and balancing the trade-offs between the accuracy and fairness of the recommendations.

Simulating the various algorithms used by this \cite{liang2016causal} paper recommendation system requires capturing complex behaviors and dynamics, such as the user discovery process, the user preference function, the causal effects of the recommendations, and the feedback loop between the users and the items.

Overall, simulating the various algorithms used by social media recommender systems requires capturing intricate behaviors and dynamics, which is a complex task that will require considerable engineering effort.

\section{Relevance and Future Work}
With the advent of regulation in the sphere of social media and digital services such as the Digital Services Act, the Online Safety Bill, and other legislation, global lawmakers are hoping to strike a balance and deliver effective policy mechanisms to improve the online experience for users of social platforms. 
This is challenging in the absence of access to platform data, knowledge of business considerations, and a lack of clarity into what is called 'impossible tradeoffs' in the field of trust and safety--which often balances availability of resources against provision of additional mechanisms to ensure user safety.
Simulators offer a high-fidelity solution that trades off complexity from real-world algorithms with the intuitive understanding they offer to non-domain experts, about the workings of a complex system such as social networks bridging the public-private information gaps to effectively address online safety issues.

\section{Conclusion}

Multiagent simulators are expressive models of online interaction and have demonstrably yielded value in varied applications.
While there are limitations from scale and complexity, there is significant value that is likely to be unlocked by advances in computational modeling and machine learning for this area.

\bibliography{pmlr-sample}

\appendix

\end{document}